\documentclass[epj]{webofc}
\usepackage[varg]{txfonts}   
\usepackage{amsmath}
\wocname{EPJ Web of Conferences}
\woctitle{CONF12}
\usepackage{hyperref}

\def\hri#1#2{\href{http://arxiv.org/abs/#1}{[arXiv:#1]#2}}
\def\hre#1#2{\href{http://arxiv.org/abs/#1/#2}{[arXiv:#1/#2]}}

\def\be{\begin{equation}}
\def\ee{\end{equation}}
\newcommand{\bea}{\begin{eqnarray}}
\newcommand{\eea}{\end{eqnarray}}
\def\l{\lambda}

\def\la{\lambda}
\def\m{\mu}
\def\n{\nu}

\def\t{\tau}

\def\order#1{$\mathcal{O}\left(#1\right)$}
\def\morder#1{\mathcal{O}\left(#1\right)}
\def\Awf{{A}} 
\def\ag{\mathfrak{a}}

\def\ax{\mathfrak{a}}

\begin{document}
\selectlanguage{english}
\title{Theta angle in holographic QCD}
%
%

\author{Matti J\"arvinen\inst{1}\fnsep\thanks{\email{jarvinen@lpt.ens.fr}} 
}

\institute{
Laboratoire de Physique Th\'eorique de l' \'Ecole Normale Sup\'erieure
\& Institut de Physique Th\'eorique Philippe Meyer, 
PSL Research University,  CNRS,  Sorbonne Universit\'es, UPMC
Univ.\,Paris 06, 
24 rue Lhomond, 75231 Paris Cedex 05, France
}

\abstract{%
V-QCD is a class of effective holographic models for QCD which fully includes the backreaction of quarks to gluon dynamics. The physics of the $\theta$-angle and the axial anomaly can be consistently included in these models. We analyze their phase diagrams over ranges of values of the quark mass, $N_f/N_c$, and $\theta$, computing observables such as the topological susceptibility and the meson masses. At small quark mass, where effective chiral Lagrangians are reliable, they agree with the predictions of V-QCD. 
}
\maketitle
\section{Introduction}
\label{intro}

Most of the literature on holographic QCD assumes the 't~Hooft limit,
\be
N_c \to \infty \,,\qquad \lambda= g_{YM}^2 N_c =\mathrm{fixed} \,,
\label{tHl}\ee
and keeping also the number of flavors $N_f$ fixed. In this limit, the quarks act as probes in the gluon dynamics. This simplifies the analysis drastically, but keeps many important features of the theory. For example, a meson spectrum very similar to that observed in experiments can be obtained. 

There are, however, many features related to the backreaction of the quarks to the gluons that cannot be accessed in the 't~Hooft limit. In particular, the conformal window in QCD is found at finite values of the ratio $N_f/N_c\equiv x$, whereas in the 't~Hooft limit this ratio tends to zero. Also ordinary QCD has two or three light quarks (depending on whether the strange quark is included or not) so the ratio of quark and gluon degrees of freedom $\sim N_f/N_c$ is not small and there is sizeable backreaction. In order to study backreaction in holographic QCD, we will be working in another large $N$ limit, the Veneziano limit~\cite{vu1}:
\be
N_c \to \infty \,,\qquad N_f \to \infty \,,\qquad {N_f \over N_c}=x=\mathrm{fixed} \,,\qquad \lambda= g_{YM}^2 N_c =\mathrm{fixed} \,.
\label{vl}\ee

This talk will concentrate on backreacted holographic bottom-up models for QCD, which were introduced in~\cite{jk}. These models were coined V-QCD, where the letter V refers to the Veneziano limit. The idea is to follow principles from string theory and top-down constructions closely, but allow for shortcuts whenever derivation from first principles is either not possible or turns out to be too involved. This leads to a class of effective holographic models, roughly analogous to effective field theory, which are more advanced than the typical bottom-up models in the literature. The action of V-QCD contains several potentials which need to be determined by comparing to QCD data obtained from experiments, lattice studies, or perturbation theory. Such comparison leads to a unified model for SU$(N_c)$ gauge theories at various values of $x$, the quark mass, and $\theta$-angle, which also can be studied at finite temperature, quark chemical potential, and external magnetic field.

The V-QCD models are constructed from two building blocks. The first block is improved holographic QCD (IHQCD) which is a string-inspired model for pure Yang-Mills theory using defined in terms of a properly adjusted five dimensional dilaton gravity~\cite{ihqcd,ihqcd2} (see also~\cite{Gubser:2008ny}).  
The second building block is a prescription for adding flavor through a Sen-like tachyon Dirac-Born-Infeld (DBI) action, which describe the dynamics of a pair of space filling $D4-\overline{D4}$ branes~\cite{bcckp,ckp}. 

Putting together the two building blocks in the Veneziano limit, where the flavor fully backreacts to the glue, defines the V-QCD models~\cite{jk}.  Due to the backreaction the model is also rich and complicated, and it has to be solved numerically, but the results agree well with the expected physics of QCD. In particular, the backreaction makes it possible to study holographically the physics of the conformal window and the conformal transition~\cite{jk,Jarvinen:2009fe} (see also~\cite{son,Alanen:2010tg,kutasov,kutasovdbi,parnachevdbi,Alvares:2012kr,alhoevans,scott}).

The properties of V-QCD, among other things the phase diagram at finite temperature and chemical potential~\cite{alho,Alho:2013hsa,Alho:2015zua}, the meson spectra~\cite{letter,Arean:2013tja,Iatrakis:2014txa}, and the phase diagram at finite quark mass~\cite{Jarvinen:2015ofa}, have already been studied extensively, and the model has been found to give a nice overall description of QCD. Also physics of the heavy ion collisions, including direct photon production~\cite{Iatrakis:2016ugz} and effects due to finite background magnetic field~\cite{Drwenski:2015sha,GIJN}, are being studied. For a brief, generic introduction to V-QCD, see~\cite{Jarvinen:2015wia}.
In this talk V-QCD and its application to the physics of axial anomaly and the $\theta$-angle~\cite{Arean:2016hcs} will be reviewed. CP-odd physics as also been analyzed using top-down holography~\cite{Barbon:2004dq,Armoni:2004dc}, in particular in great detail in the Witten-Sakai-Sugimoto model recently~\cite{Sakai:2004cn,Bigazzi:2014qsa,Bigazzi:2015bna,Bartolini:2016dbk}. 

As is well known, the $U(1)_A$ part of the flavor symmetry in QCD is broken by quantum effects, leading to the absence of a light pseudo Goldstone mode in the spectrum: instead, a large mass of $\eta'$, has been observed experimentally. Nontrivial topological gauge field configurations,
instantons,  violate the $U(1)_A$ symmetry~\cite{'tHooft:1976up}.
The physics of the anomalous breaking of $U(1)_A$ is tricky in the limit of large $N_c$, where one would naively expect the instanton contributions to be highly suppressed. This turns out not to be the case, however, since the instanton gas ``disappears'' at large $N_c$~\cite{Witten1}, and moreover 
the resulting $U(1)_A$ breaking takes place at leading order in $1/N_c$ in the Veneziano limit~\cite{vu1}. The effect of the axial anomaly can be controlled at small $x$, where the mass of the $\eta'$ meson is related to the Goldstone modes (the pions) by the Witten-Veneziano relation,
\be \label{WVrel}
 m_{\eta'}^2 \simeq  m_\pi^2 + \frac{N_f}{N_c} \ \frac{\chi}{\bar f_\pi^2} \,,
\ee
where $\chi$ is the Yang-Mills topological susceptibility and $\bar f_\pi$ is the pion decay constant, which is normalized such that $\bar f_\pi = \morder{N^0}$.

The instanton vacua give rise to an additional periodic parameter, the $\theta$-angle, which is identified as the coupling of the CP-odd ${\mathbb T}\mathrm{r}\,[G \wedge G]$ operator in QCD.  After coupling to quarks the physical, gauge invariant parameter is the combination $\bar\theta =\theta + \sum_a \phi_a$, where $\phi_a$ are the phases in the masses of the various quark flavors.

In this talk the V-QCD models at zero $\bar\theta$ will be reviewed first. Then the implementation of $\bar\theta$ in V-QCD, the CP-odd dynamics, and results for related observables will be discussed.

\section{V-QCD at $\bar\theta=0$}

\subsection{Glue sector: IHQCD}

This first building block of V-QCD is the model for the gluon sector, IHQCD, which is based on five dimensional Einstein-dilaton gravity~\cite{ihqcd,ihqcd2}. 
We use a Poincar\'e covariant Ansatz for the metric
\be
ds^2=e^{2 \Awf(r)} (dx_{1,3}^2+dr^2)\,.
\label{bame}
\ee
The dictionary for the glue sector is:
\begin{itemize}
 \item The exponential of the dilaton $\l=e^\phi$ is dual to the operator ${\mathbb T}\mathrm{r}\, [G^2]$, and its background value is identified with the 't Hooft coupling.
 \item The warp factor $A$ is mapped to the logarithm of the energy scale in field theory.
 \item The metric is dual to the energy-momentum tensor.
\end{itemize}
Explicitly, the glue action takes the form
\be
S_g= M^3 N_c^2 \int d^5x \ \sqrt{-g}\left(R-{4\over3}{
(\partial\lambda)^2\over\lambda^2}+V_g(\lambda)\right) \,,
\label{vg}
\ee
where the dilaton potential $V_g$ is roughly mapped to the $\beta$-function of Yang-Mills.

IHQCD was compared to experimental data (and lattice data for small $N_c$) in~\cite{ihqcdt,ihqcdt2,data}, and good agreement with the glueball spectrum and thermodynamics could be obtained. The equation of state of the model agrees almost perfectly with the large $N_c$ lattice data~\cite{Panero:2009tv} which was computed afterwards.

\subsection{Flavor sector: Tachyonic DBI action}

The flavor sector contains the following (CP-even) fields:
\begin{itemize}
 \item The tachyon $\t$ which is dual to the $\bar \psi\psi$ operator and therefore sources the quark mass. We will assume here that the quark mass is flavor independent.
 \item The gauge fields $A_{L/R}$ which are dual to the left- and right-handed currents $\bar \psi (1\pm \gamma_5) \gamma_\mu \psi $. 
\end{itemize}
The dynamics  is given by the tachyon DBI action~\cite{ckp}
\be \label{vf}
 S_f =- {N_fN_c} M^3 \int d^5 x\, {V_f}(\la,\tau) \sqrt{-\det(g_{ab}+ {\kappa}(\la) \partial_a \tau \partial_b \tau)} \,,
\ee
where we only included the terms and structure relevant for the vacua at $\bar\theta = 0$. In particular, the gauge fields were set to zero. The tachyon potential is assumed to have ``Sen-like'' dependence on $\t$,
\be
{V_f}(\la,\tau) = V_{f0}(\l) \exp(- a(\l) \t^2)\,.
\ee
For chirally broken solutions the tachyon runs to infinity in the IR so that the tachyon potential vanishes, corresponding to the annihilation of the $D4-\overline{D4}$ pair deep in the IR. 
This action has been successfully tested in the probe limit~\cite{ikp1,ikp2}: tachyon condensation in the bulk, triggered by confinement, was seen to produce the desired picture of chiral symmetry breaking and the spectrum of light mesons was very close to that observed in experiments.

\subsection{The fusion: V-QCD}

The action of the V-QCD models reads simply $S_{V-QCD}  = S_g + S_f$,
where both terms are \order{N^2} in the Veneziano limit so that flavor is fully backreacted to the glue.

Because we are working with an effective model, we need to specify the functions $V_g(\l)$, $V_{f0}(\l)$, $a(\l)$, and $\kappa(\l)$ of~\eqref{vg} and~\eqref{vf} in order to fully pin down the model. As it turns out, there are several qualitative constraints both in the UV ($\l \to 0$) and in the IR ($\l \to \infty$). 

In the UV, i.e., at weak coupling, holographic description through classical gravity is not expected to be accurate because neglected higher curvature corrections would become important. Therefore we determine the UV behavior by relying on perturbation theory for QCD, which is expected to provide the best possible UV ``boundary conditions'' for the more interesting IR physics. 
That is, we match the potentials $V_g$, $V_{f0}$ and $\kappa/a$  with the $\beta$-function and the anomalous dimension of the quark mass in QCD perturbatively in $\l$~\cite{ihqcd,ihqcd2,jk}.

In the IR, there are several constraints which are obtained from the regularity of the solutions as well as by comparing to qualitative features of QCD, such as confinement, asymptotics of the spectra, and behavior at large quark mass~\cite{jk,Arean:2013tja,Jarvinen:2015ofa}. These constraints single out certain power law asymptotics for the various functions as $\l \to \infty$. Interestingly, best results are obtained with exactly the same powers as one would have expected by using arguments from string theory, possibly up to logarithmic corrections in $\l$.

In the intermediate regime, i.e., for $\l =\morder{1}$, the above constraints do not restrict the potentials, and the potentials should be fitted quantitatively to QCD data from experiment and the lattice.

\section{CP-odd physics in V-QCD}

\subsection{CP-odd action and its symmetries}

The CP-odd fields are
\begin{itemize}
 \item The axion $\ag$ which is dual to the operator $\mathbb{T}\mathrm{r}\, G \wedge G$ and sources
 the $\theta$-angle on the boundary.
 \item The phase of the tachyon $\xi$ which is dual to $\bar \psi\gamma_5 \psi$ and
 sources the phase of the quark mass. 
\end{itemize}
For the mechanism for implementing the CP-odd terms we follow~\cite{bcckp,ckp}. The mechanism was adapted to the backreacted case in~\cite{Arean:2013tja,Arean:2016hcs}.

The anomalous $U(1)_A$ symmetry is implemented through the following transformations in the bulk:
\be \label{symm}
\xi\to \xi-2\epsilon\ , \qquad \mathfrak{a}\to \mathfrak{a}+2x\,V_a\,\epsilon\ , \qquad A_{\m}\to A_{\m}+\partial_{\m}\epsilon \,,
\ee
where $A_\m$ is the bulk $U(1)_A$ gauge field. This maps to the correct anomaly
\be
\partial_{\mu} J^{(5)\,\m} = {N_f \over 16 \pi^2}\, \epsilon^{\m\n\rho\sigma}\,
{\mathbb T}\mathrm{r}\, (G_{\m\n}\, G_{\rho\sigma})+2i\, m_q \,\bar\psi^i\gamma^5\psi^i
\ee
for QCD if the potential $V_a \to 1$ at the boundary. The invariant field
$\bar \ax = \ax + x\, \xi\, V_a $
sources the $U(1)_A$ invariant combination $ \bar\theta/N_c = \theta/N_c +\arg(\det M_q)/N_c$, where $M_q$ is the quark mass matrix. If $m_q=0$, $\bar\theta$ is not well defined, and in this case the $\theta$-angle can be gauged away as is the case in QCD. 

The full action for V-QCD, including the CP-odd terms, reads
\be
  S_{V-QCD}  = S_g + S_f + S_a
\ee
where the last term
\be \label{Sadef}
 S_a = -{M^3\,N_c^2\over2}\int d^5x\, \sqrt{-\det g}\,Z(\lambda)\left[d\mathfrak{a}-x\left(2V_a\,A-\xi\, dV_a \right)\right]^2
\ee
arises from the Wess-Zumino-Witten action for the $D4$ branes and is covariant under~\eqref{symm}.

Again to fully determine the model, we also need to specify the potentials $V_a(\l,\t)$ and $Z(\l)$. The latter has already been analyzed in the context of CP-odd physics of Yang-Mills theory for the IHQCD model~\cite{ihqcd,ihqcd2,data}. The potential $V_a$ plays a crucial role in the coupling to flavors. Because the axial anomaly should not receive perturbative corrections, $V_a$ should not have a series expansion in $\l$ near the boundary. A natural way to avoid this is to assume that $V_a$ depends on the tachyon $\t$ only, and take $V_a = \exp(- b \t^2)$.  The IR regularity of the solutions requires that $V_a$ vanishes faster than $V_f$ at large $\t$.

Since $\bar\theta$ is an angle, the physics should be unchanged if it is shifted by $2\pi$. This is not immediately obvious in our model: $\bar\theta$ is linked to the phase of the tachyon, but the dictionary implies that a $2\pi$ rotation of the phase shifts $\bar\theta$ by $2\pi N_f$ instead of $2 \pi$. The expected periodicity is, however, built in the model 
and appears through branch structure of its vacua~\cite{Arean:2016hcs}, 
as expected for QCD at large $N_c$~\cite{Witten:1998uka}.

\begin{figure}[ht]
\centering
\includegraphics[width=7cm,clip]{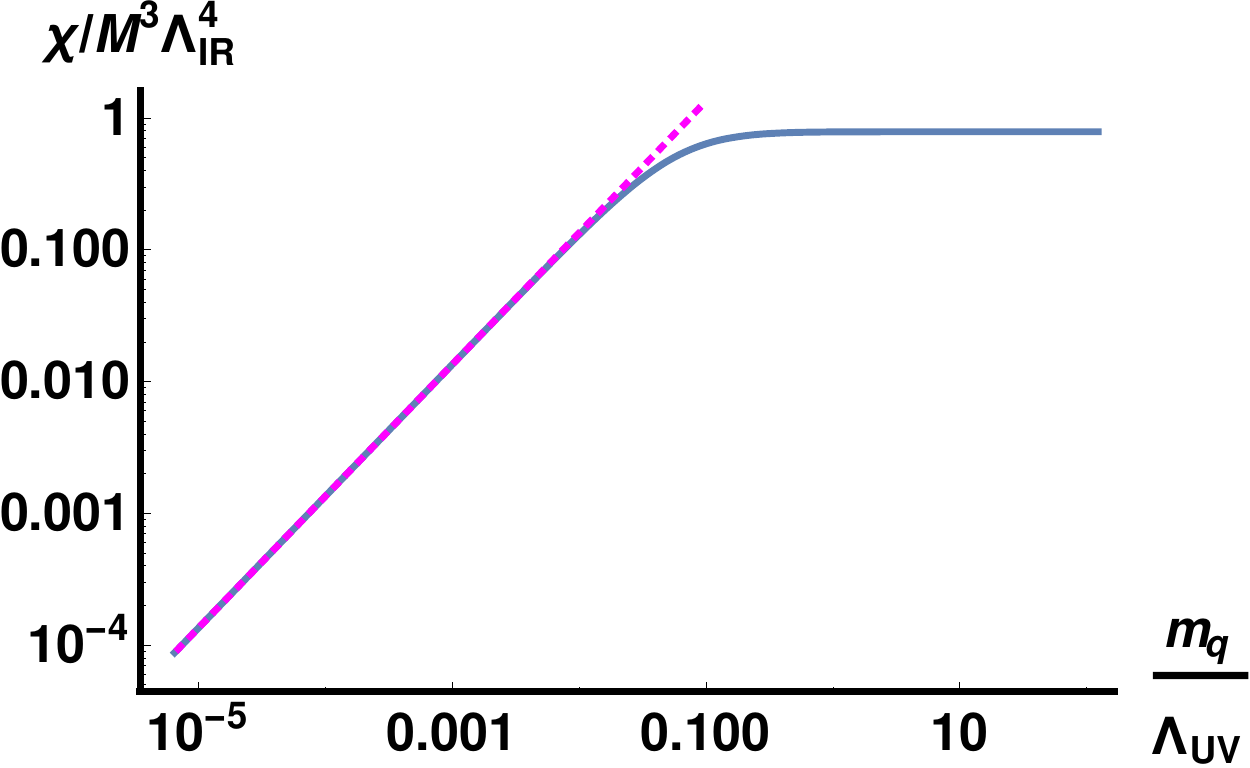}
\caption{The dependence of the topological susceptibility on the quark mass in the ``standard'', dominant vacuum at $x=2/3$ and $\bar\theta = 0$.
 The blue solid curve is the results for V-QCD and the dashed magenta curve is the prediction from chiral Lagrangians at small $m_q$.}
\label{fig-1}      
\end{figure}

\subsection{Free energy and topological susceptibility}

We will now present some (mostly numerical) results for the model, starting from the free energy and topological susceptibility, and choosing a value of $x=2/3$ which is corresponds to regular QCD with two light quark flavors. For the numerical results we chose a set of potentials (the potentials~I in~\cite{Arean:2013tja,Arean:2016hcs}), which satisfy the qualitative constraints discussed above but have not yet been matched quantitatively to QCD data. As we shall see, the holographic model agrees perfectly with chiral Lagrangians in the regime of small quark mass, where the effective field theory approach is reliable. We stress, however, that V-QCD is more general, producing a model for the full $m_q$ dependence as these observables, for example. Similar results as those presented below have also been found recently in the Witten-Sakai-Sugimoto model~\cite{Bigazzi:2014qsa,Bigazzi:2015bna,Bartolini:2016dbk}.

The topological susceptibility $\chi$ can be defined as the second derivative of the free energy with respect to $\bar\theta$. We show the result for V-QCD as a function of the quark mass\footnote{The UV and IR scales $\Lambda_\mathrm{UV}$ and $\Lambda_\mathrm{IR}$ can be defined as the characteristic scales of the UV and IR expansions of the metric and the dilaton.} in Fig.~\ref{fig-1}. For small $m_q$, we recover the expected result (see, e.g.,~\cite{Leutwyler:1992yt})
\be
 \chi \simeq -\frac{m_q \langle \bar\psi\psi\rangle}{N_f^2} \,,
\ee
and $\chi$ approaches the constant value at large $m_q$. This is also expected because the quarks are effectively decoupled as $m_q \to \infty$ and $\chi$ is determined solely through gluon dynamics.

\begin{figure}[ht]
\centering
\includegraphics[width=6.7cm,clip]{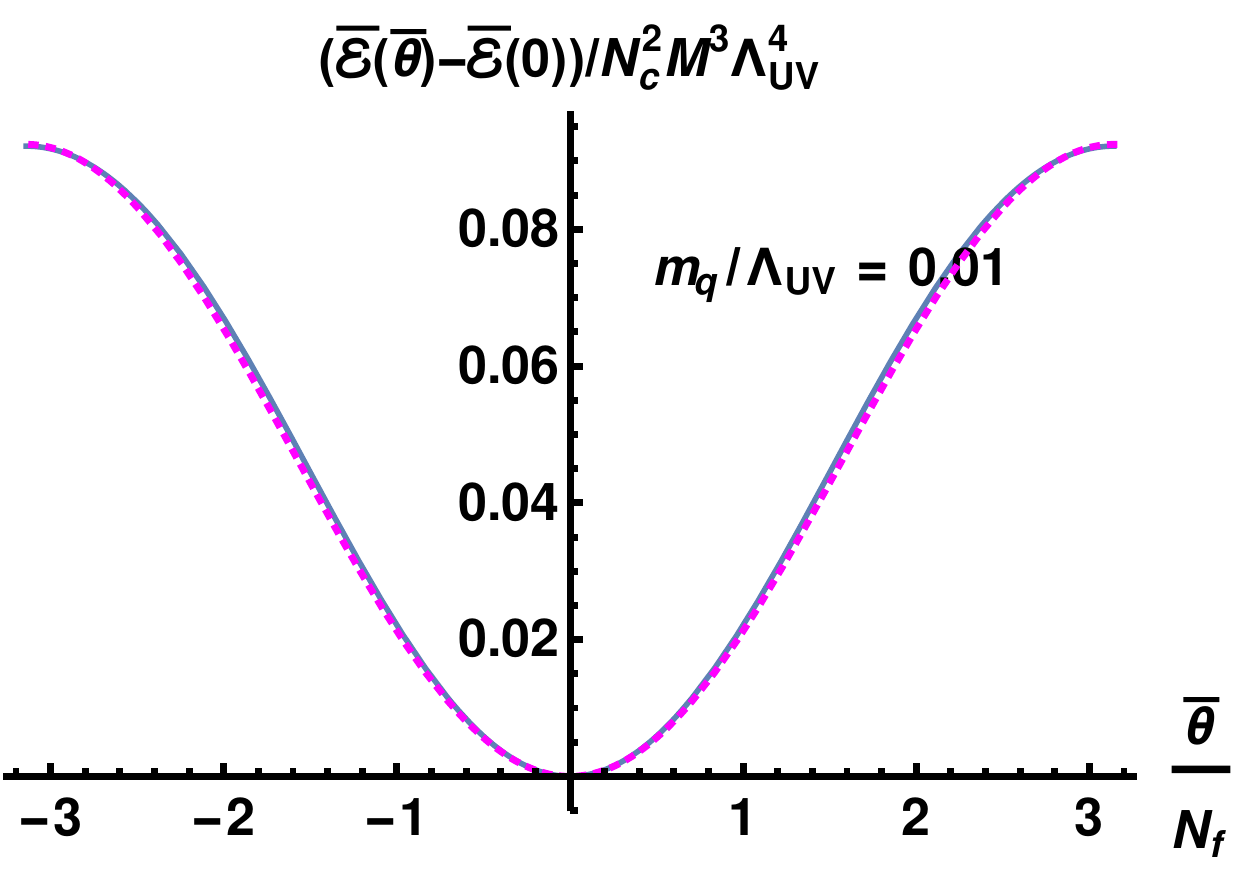}%
\hspace{2mm}\includegraphics[width=6.7cm,clip]{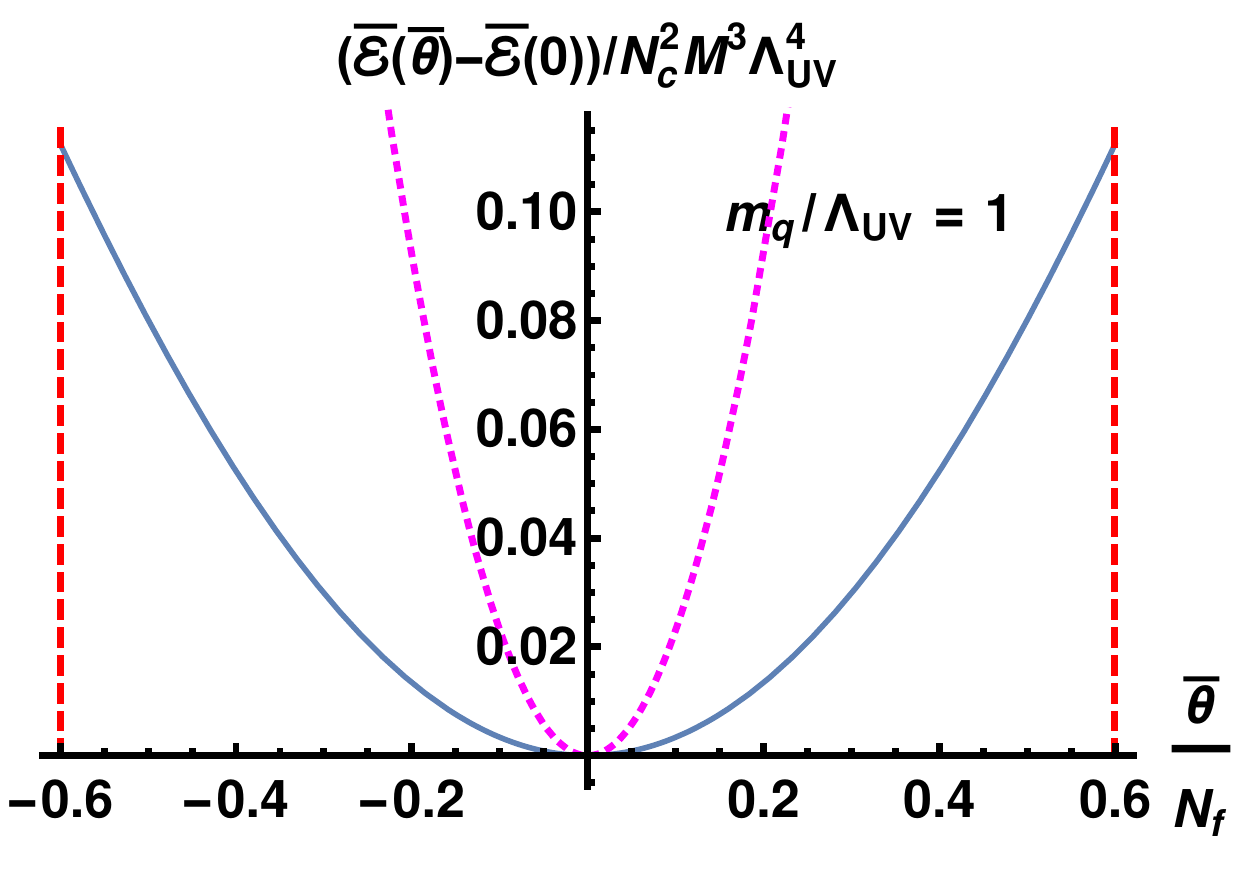}
\caption{The dependence of the free energy on the $\bar \theta$-angle at $x=2/3$ for $m_q=0.01$ (left plot) and $m_q=1$ (right plot). The dashed magenta curves are again predictions from chiral Lagrangians. The dashed red vertical lines denote where the solution ceases to exist for $m_q=1$.}
\label{fig-2}    
\end{figure}

We have also studied the $\bar\theta$ dependence of the free energy, which is shown in Fig.~\ref{fig-2} for two values of the quark mass. These plots are for a single branch of vacua and are not $2\pi$ periodic in $\bar\theta$ -- as pointed out above, other branches will need to be added to recover this periodicity.\footnote{As a consequence, the curves of Fig.~\ref{fig-2} will correspond to the physical vacuum only very close to $\bar\theta = 0$.} For $m_q=0.01$ (left plot), V-QCD is close to the result of the chiral limit,
\be \label{Echiral}
 \mathcal{E} = -m_q\langle\bar\psi\psi\rangle\left(1-\cos (\bar\theta/N_f)\right) \,,
\ee
given as the dotted magenta curve, but at larger $m_q$ (right plot) there is already a large difference between the full numerical result and~\eqref{Echiral}.

The result~\eqref{Echiral} arises from an effective Lagrangian containing the pion fields, but no $\eta'$ meson. For small $x$ the $\eta'$ meson is almost as light as the pions, and including it in the Lagrangian leads to~\cite{Leutwyler:1992yt,Witten:1980sp}
\be
 \mathcal{E} = \min_{\xi_0}[-m_q\langle\bar\psi\psi\rangle\left(1-\cos \xi_0\right) +\chi_\mathrm{YM}(N_f \xi_0-\bar\theta)^2/2] \,.
\ee
It can be shown that, interestingly, V-QCD produces this result as well (for small $m_q$ and $x$)~\cite{Arean:2016hcs}. After taking into account the full branch structure, the free energy for the dominant vacuum becomes
\be
\mathcal{E}(\bar\theta) = \frac{1}{2}\chi \min_k \left(\bar\theta+2\pi k\right)^2 \,.
\ee

\begin{figure}[ht]
\centering
\includegraphics[width=6.7cm,clip]{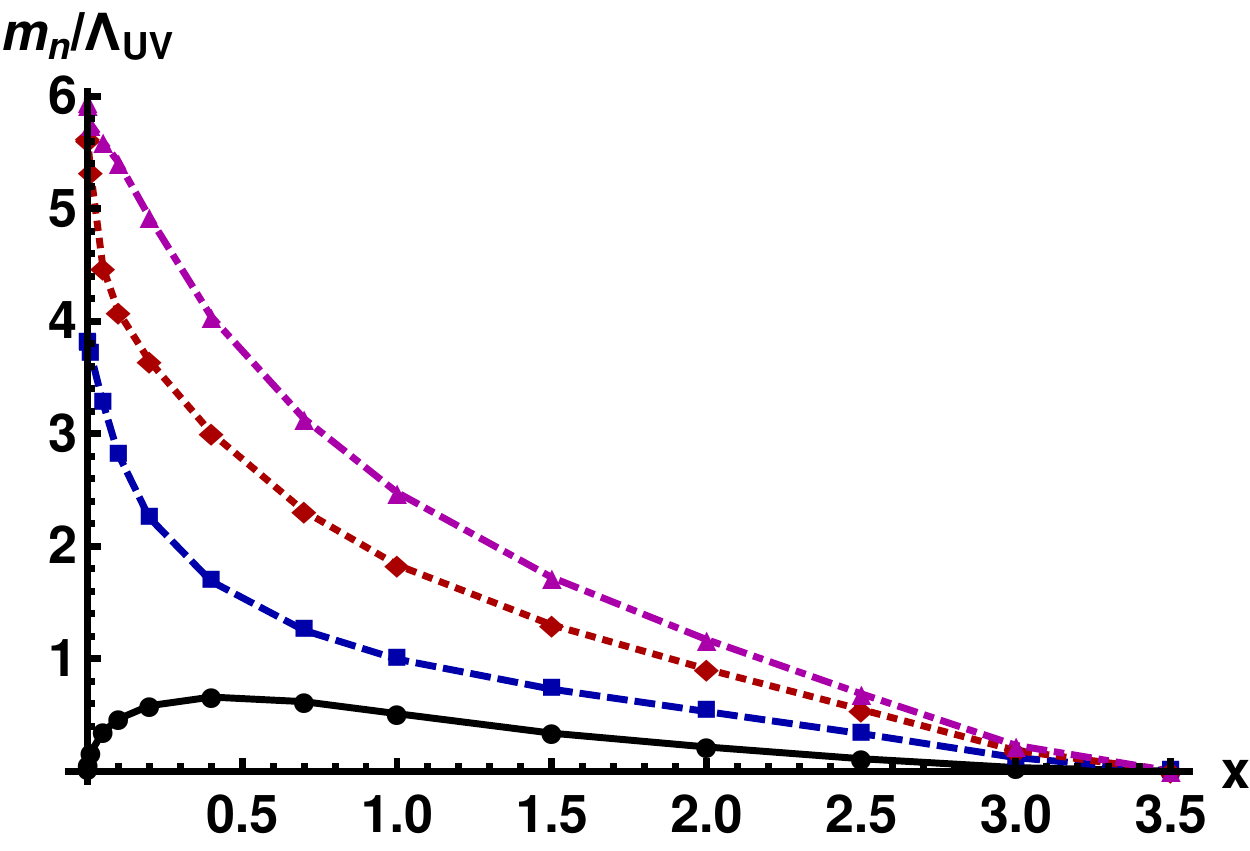}%
\hspace{2mm}\includegraphics[width=6.7cm,clip]{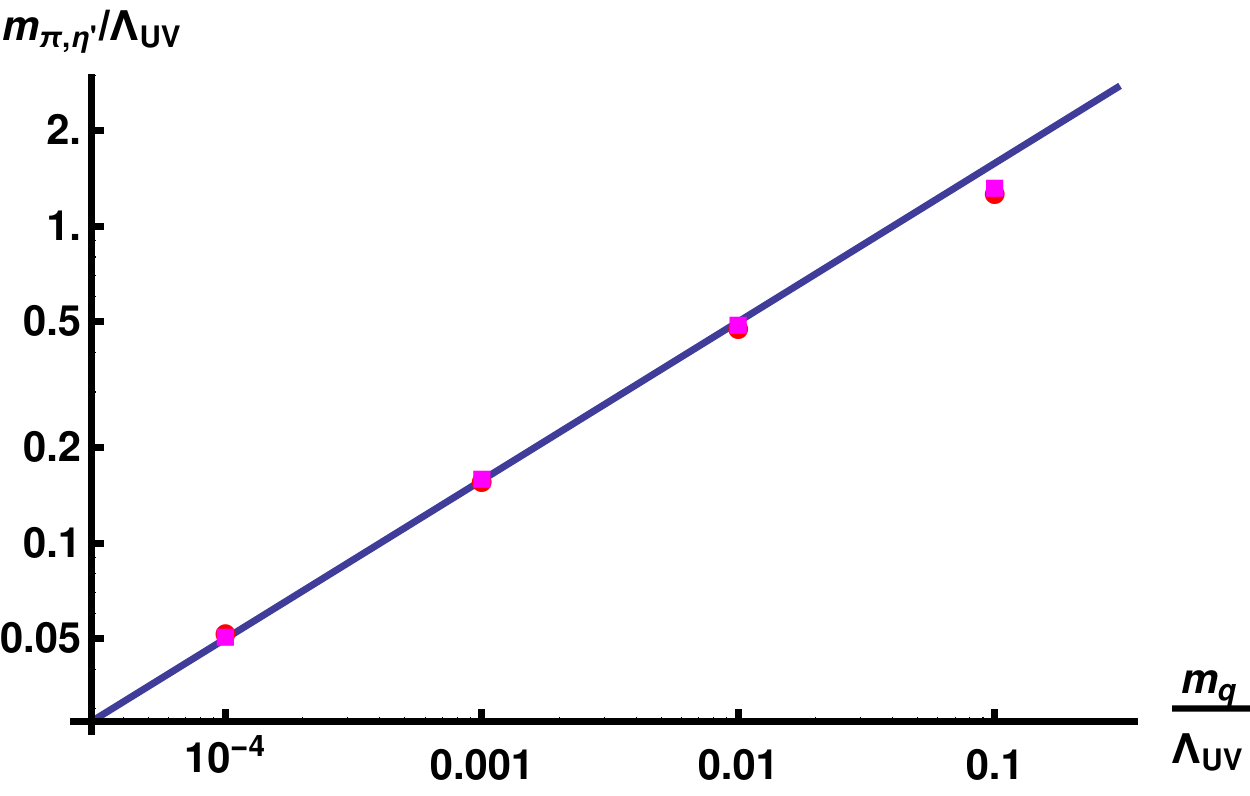}
\caption{Left: masses of the lowest four singlet pseudoscalar states as a function of $x$. Right: the dependence of the mass of the $\eta'$ (red circles) and the pion mass (magenta squares) on $m_q$ at $x=0.0001$. The blue line is a fit to the Gell-Mann-Oakes-Renner relation.}
\label{fig-3}   
\end{figure}

\subsection{CP-odd meson spectrum at $\bar\theta=0$}

The spectrum of $0^{-+}$ mesons contains glueballs and (flavor singlet) $\bar\psi\psi$ states which are mixed for finite $x$. We plot the masses of the four lowest states as a function of $x$ (at $\bar\theta=0=m_q$) in Fig.~\ref{fig-3} (left). The lowest state is identified as the $\eta'$ meson as $x \to 0$. It can be checked, numerically and analytically, that its mass satisfies the Witten-Veneziano relation in~\eqref{WVrel} even after taking account the mixing with the pseudoscalar glueballs. Numerical verification at small $x$ is shown in the right hand plot in Fig.~\eqref{fig-3}, where we plot the masses of the pions and the $\eta'$ as a function of $m_q$ at $x=0.0001$. As expected from the Witten-Veneziano and Gell-Mann-Oakes-Renner relations, the data points are on top of each other, and the masses vanish as $\sqrt{m_q}$ as $m_q \to 0$.

\begin{figure}[ht]
\centering
\includegraphics[width=9.5cm,clip]{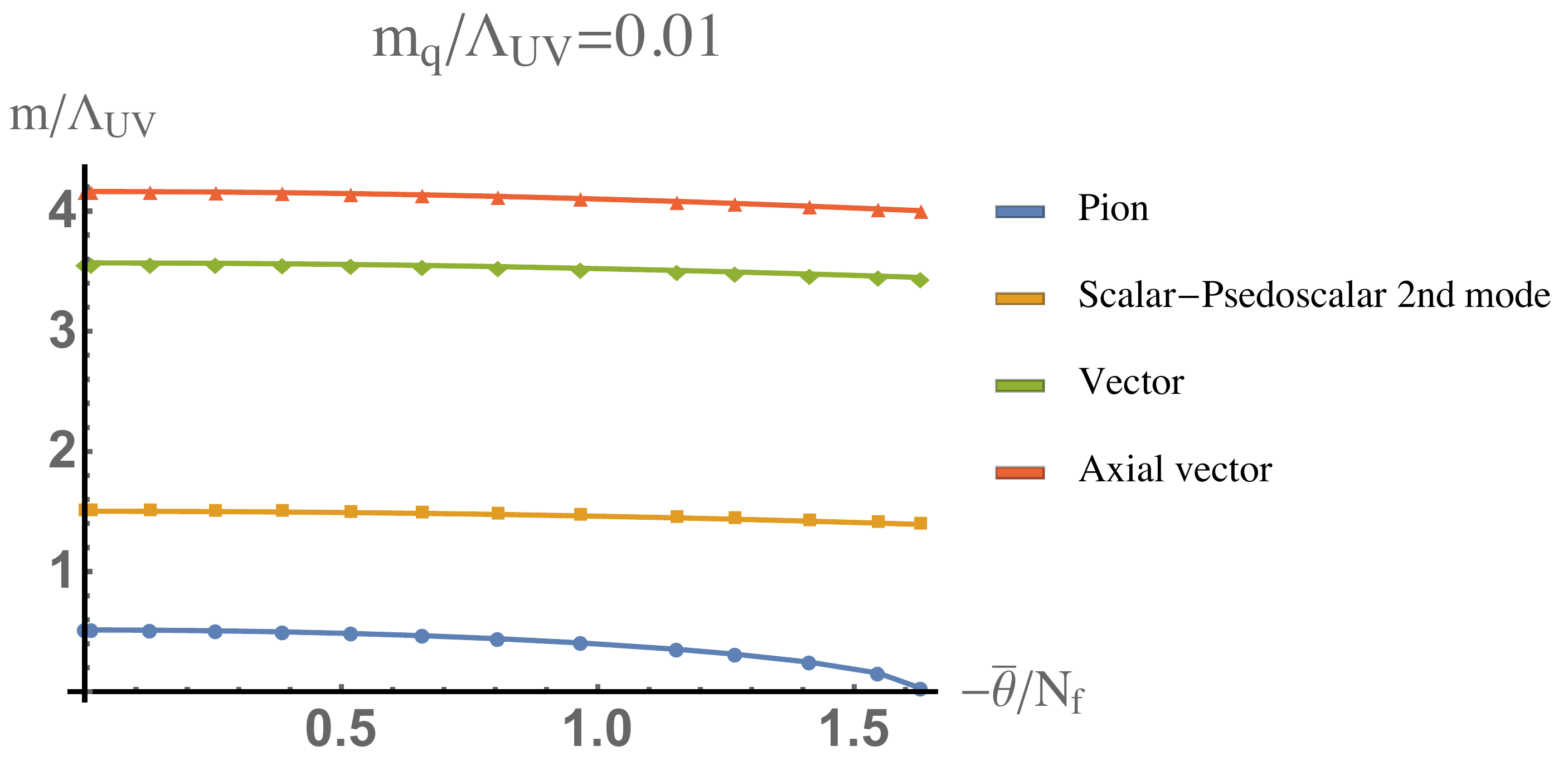}
\caption{The masses of the lowest (flavor nonsinglet) meson states as a function of $\bar\theta$ at $x=2/3$ and $m_q/\Lambda_\mathrm{UV}=0.01$. }
\label{fig-4}   
\end{figure}

\subsection{Meson spectrum at $\bar\theta\ne 0$}

We have also studied the meson spectrum at $\bar\theta\ne 0$, restricting to the flavor nonsinglet states which obey simpler fluctuation equations than the singlet states. Fig.~\ref{fig-4} shows the results for the masses of the lowest states in each meson sector at $m_q=0.01$ for a single branch of vacua as a function of $\bar\theta$. The pseudoscalars and scalars mix for $\bar\theta \ne 0$. The masses of other states than the pion are only weakly dependent on $\bar\theta$. The pion mass is close to the result of the chiral limit, which generalizes the Gell-Mann-Oakes-Renner relation to finite $\bar\theta$,
\be
 f_{\pi}^2\, m_\pi^2 =  -\langle \bar \psi\psi\rangle\big|_{m_q=0}\, m_q\, \cos \frac{\bar\theta}{N_f} +\morder{m_q^2} \,.
\ee
An perturbative instability therefore sets in at $|\bar\theta| \simeq \pi N_f$ as the squared pion mass becomes negative. Notice, however, that the branch of vacuum for which we plot the masses is dominant only for $|\bar\theta| \lesssim \pi$, which suggest that there is a long range in $\bar\theta$ for which it is metastable.

\subsection{Conformal window}

Within a range $x_c<x<11/2$, and zero quark mass, QCD flows to a nontrivial fixed point in the IR. The location $x=x_c$ of the conformal transition is nontrivial and expected to be close to 3 or 4. It is interesting to perturb the theory in the conformal window by adding a tiny quark mass. As a response to this perturbation, the meson spectrum becomes discrete and the meson masses $m_n$ obey characteristic scaling laws,
\be
 m_n \propto m_q^\frac{1}{1+\gamma_*}\,,\qquad (m_q \to 0)\,,
\ee
where $\gamma_*$ is the anomalous dimension of the quark mass at the IR fixed point. Such scaling laws are called hyperscaling relations~\cite{DelDebbio:2010ze}, and also hold in holographic models~\cite{scott,Jarvinen:2015ofa}, including V-QCD. We have derived the hyperscaling relation for the topological susceptibility from V-QCD:
\be
 \chi \propto m_q^\frac{4}{1+\gamma_*}\,,\qquad (m_q \to 0)\,.
\ee

\section{Conclusion}

V-QCD is a realistic, backreacted effective holographic model for QCD. We have tested the implementation of CP-odd physics, i.e., the $\theta$-angle and axial anomaly, in V-QCD. In the regime of small quark mass, where predictions of chiral Lagrangians are reliable, we found perfect agreement between them and the holographic model. In particular, the Witten-Veneziano and generalized Gell-Mann-Oakes-Renner relations were verified. In the opposite limit $m_q \to \infty$, quarks are decoupled, leaving us with the results for pure Yang-Mills theory. We also derived the hyperscaling relation for the topological susceptibility in the conformal window.

In an ongoing work V-QCD is fitted to experimental and lattice data (such as meson masses and the equation of state at finite temperature), which will constrain the dependence of the various potentials of the V-QCD action at intermediate $\l=\morder{1}$. The aim is to produce an overall effective description of QCD in the Veneziano limit, which is trustable for large range of $x$, $m_q$, and $\bar\theta$, including phases with different IR dynamics, i.e., the conformal window and a phase with dynamics similar to regular QCD. Such a model will be useful for interpolating and extrapolating results to regimes where field theory computations are challenging.

\end{document}